\documentclass[linenumbers]{aastex631}

\graphicspath{{./}{figures/}}
\begin{document}

\title{3-D exact analytical solutions of two fluid plasma, MHD and neutral fluid equations for the creation of ordered structures as well as jet-like flows}

\author{Hamid Saleem}
\affiliation{Theoretical Research Institute \\ Pakistan Academy of Sciences, 3-Constitution Avenue, G-5/2 \\ Islamabad 44000, Pakistan.}

\author{Zain H. Saleem}
\affiliation{Argonne National Laboratory \\
9700 S. Cass Ave  \\
Lemont, IL 60439, USA.}



\begin{abstract}

The 3-D exact analytical solutions of ideal two fluid plasma, single fluid plasma (MHD) and neutral fluid equations have been found using physically justifiable assumptions. Surprisingly these solutions satisfy all non-linearities in the systems. It is pointed out that these solutions explain the fundamental mechanism behind the creation of vast variety of ordered structures in plasmas and fluids. In the limiting case of two dimensional (2-D) dependence of fields, the theoretical model for plasma is applied to explain the formation of spicules in solar chromosphere. It is pointed out that the main contribution of electron (ion) baro clinic vectors is to produce vorticity in the plasma and that magnetic field generation is coupled with the flow of both electrons and ions. 

\end{abstract}

\keywords{Solar Coronal Heating, Creation of Ordered Structures, Jet-like Flows}


\section{\label{sec:level1}Introduction}

Ordered structures appear in both laboratory \cite{dnestrovskij2005self,kugland2012self} and astrophysical \cite{zweibel1997magnetic, widrow2002origin, ryu2008turbulence} plasmas. Self-organization is a natural process for producing global structures on macroscopic scale as the complex system evolves from microscopic turbulent state \cite{yoshida2002variational, yoshida2014self}. Astrophysical jets were observed long ago \cite{curtis1918descriptions} and later several similar observations were reported in astrophysical environments such as the material outflows from young stellar objects (YSOs), accretion disks, and active galactic nuclei (AGN) \cite{jennison1953fine, mestel1961note, ferrari1998modeling}. Solar spicules, coronal loops, coronal holes, surges, prominences, flares are examples of ordered structures emerging from highly nonlinear solar plasma dynamics \cite{priest1982solar, woo1996kilometre, aschwanden1999coronal, de2007chromospheric, aschwanden2017width, klimchuk2020cross}.
Material ejection in several different forms in the perpendicular direction to the solar plasma is very common in its atmosphere which is divided into three regions, surface (photosphere), chromosphere and corona. The solar photosphere is a thin layer of plasma having thickness of about $500$ km and above it lies the chromospehere with a thickness of $2500$ km. Solar atmosphere is not smooth, rather it has inhomogeneities of different scales and plasma structures of different forms. Average electron density $n_e$ in the photosphere is $3\times (10^{13}) cm^{-3}$, temperature is low, therefore the plasma contains neutral atoms and is colissional as well. In the chromosphere, electron density is in the range $3 \times (10^{10}-10^{11}) cm^{-3}$ but the temperature is higher than the photosphere and concentration of neutral atoms is negligible. Beyond the chromosphere is the solar corona which extends to infinity and there the plasma density is very small $n_e \simeq 10^{9} cm^{-3}$. In between the chromosphere and the corona lies a thin transition region of thickness $500$ km.
Temperature varies in the vertical direction in a very peculiar manner, first it
decreases from $6600$ $K$ at surface to $4400$ $K$ at a height of $500$ $km$
and then starts increasing up to $10^{4}$ $K$ in the lower chromosphere. After that, it
increases abruptly from $10^{4}$ $K$ to $(2-8)\times 10^{5}$ $K$ in the transition region and attains the value of $(1-2)\times 10^{6}$ $K$ in the lower corona  \cite{priest1982solar,slemzin2014spectroscopic,narain1990chromospheric,klimchuk2006solving,tian2017probing,tian2018transition}. In our opinion, inhomogeneous density at plasma plane and vertical variation of temperature produce thermal force to accelerate the plasma in the perpendicular direction. This phenomenon can occur in two fluid plasma, magnetohydrodynamics (MHD) and neutral fluids.
Charged particles also flow in upward direction against Earth's gravity in the ionosphere \cite{amatucci1999inhomogeneous}. Recently, a new phenomena named as 'campfires' has been observed in solar corona by the Solar Orbiter, the joint solar mission of ESA and NASA launched on 10 February 2020 \cite{antolin2021reconnection}.
Solar corona contains many different types of magnetic tubes with hot plasma
\cite{klimchuk2020cross,litwin1993structure,hara1992high,moses1997eit,schrijver1999new}. 
Abrupt coronal plasma heating is a mystery for the astrophysicists and a great deal of work in this direction has appeared in the literature (\cite{mandrini2000magnetic,xie2017plasma,hollweg1984resonant,ionson1983electrodynamic,steinolfson1993coronal,ofman1995coronal,poedts1989numerical,poedts1990efficiency,poedts1994quality,poedts1997nonlinear,vranjes2009new,vranjes2010drift,saleem2012self,saleem2021generation}. Plasma filaments in corona have different sizes
\cite{de2007chromospheric} and scale sizes of the order of $1$ $km$ in corona have been discussed \cite{woo1996kilometre}. The coronal loop structures
are highly non-uniform and consist of threads and strands \cite{woo1996kilometre,de2007chromospheric,aschwanden2017width,de2017observations,goddard2017statistical,klimchuk2020cross}.

Horizontal and vertical flows are very common in solar atmosphere \cite{brekke1997doppler,brekke1997flows,pike1997euv,qiu2000counter,schrijver1999new,zacharias2018disentangling,johnston2019fast,chae1998temperature,peter1999doppler,brekke2000observations}. Jet-like cylindrical plasma structures (spicules) flowing vertically are observed at any time in the chromosphere with diameter of $(500-1000) km$, heights of the order of $10^{4} km$ and upward velocities in the range $(20-30) km/s$. The average number of spicules observed at solar disk any time is about half a million $(0.5) \times 10^{6}$ \cite{lang2001cambridge}. The life time is $5$ to $10$ minutes, in most of the cases \cite{priest1982solar,zhang2012revisiting,kuzma2017numerical,samanta2019generation,dover2021magnetohydrodynamic}. Formation and evolution of solar spicules has been studied through numerical simulations using single fluid plasma model \cite{dover2021magnetohydrodynamic}, the magnetohydrodynamics (MHD), and two fluid plasma equations \cite{kuzma2017numerical}.

Several years ago, an effort was made to find out an exact analytical solution of two fluid plasma equations \cite{saleem2010nonequilibrium}. It was shown that if the plasma density has an exponential like profile in the xy-plane and the temperature has inhomogeneity in the yz plane, then the longitudinally uniform flow in plasma is created along with the magnetic field. The flow and magnetic fields become functions of $(x,y,t)$ coordinates. In that work, the two dimensional (2-D) analytical solution of ideal classical plasma equations was presented to explain the generation of seed magnetic field and plasma vorticity by thermal forces \cite{saleem2010nonequilibrium} using Cartesian geometry. In addition to numerical studies, it is very interesting to find out an exact three dimensional (3-D) analytical solution of two fluid plasma and MHD equations in which the flow and magnetic fields can be expressed as functions of $(x,y,z,t)$ coordinates.

No standard mathematical procedure is available to obtain an exact non trivial three dimensional (3-D) solution of two fluid plasma equations. However, one can still find out a 3-D exact analytical solution of the set of nonlinear partial differential equations keeping in view the often observed spatial dependence of density function which has approximately exponential form. The Biermann battery effect was used to explain the generation of seed magnetic field in stars \cite{biermann1950ursprung}, galaxies \cite{widrow2002origin,lazarian1992diffusion} as well as in classical laser plasmas \cite{brueckner1974laser} using one dimensional exponential density profile. In the basic mechanism, ions were assumed to be static and non-parallel density and temperature gradients of electrons were considered to be the source for the generation of magnetic field. However, computer simulations have taken into account the ion dynamics and several other effects as well. Here we present an exact 3-D solution of two fluid plasma equations to explain the fundamental mechanism behind the creation of velocity, magnetic field, and ordered structures. This solution is also applicable to single fluid plasma model (MHD) for the generation of vorticity.
 
Formation of solar spicules is also discussed using the limiting 2-D version of the two fluid plasma model. Exponential like density profile in xy-plane and linear variation of electron ion temperatures along z-axis produce upward acceleration in the plasma slab of area $xy$ and height $h$ which becomes larger than the gravitational constant $g_{\odot}$ at Sun's atmosphere, giving rise to spicules in the solar chromosphere. A more general 3-D exact solution of neutral fluid equations is presented as well.

\section{Exact 3-D Solutions of Plasma and Fluid Equations}

\subsection{Two-Fluid Plasma}
Let us consider the two fluid ideal plasma in which $\Omega_e=\frac{eB}{m_e c}$ is the electron gyro frequency, $\omega_{pe}= (\frac{4 \pi n e}{m_e})^{1/2}$ is the electron plasma oscillation frequency, and $ \lambda_{De}=(\frac{T_e}{4 \pi n e^2})^{1/2}$ is the electron Debye length.
Our aim is two-fold, first is to find out the spatial dependence of density $n$ and temperatures $T_j$ where $j=e,i$ on $(x,y,z)$ coordinates which should be the exact analytical solution of the two fluid plasma equations and it should also determine the structural forms and time evolution of flow ${\bf v}_j$ and magnetic {\bf B} fields. Second is to use 2-D density spatial profile and 1-D spatial dependence of electron (ion) temperatures $T_e (T_i)$ which give an upward acceleration to a plasma slab to produce spicules in the solar chromosphere. The latter will be discussed in section \ref{solarspicule}.
The two fluid momentum conservation equations for electrons and singly charged ions in the gravitational field ${\bf g}$ can be expressed as,
\begin{equation}
m_e n_e (\partial_t + {\bf v}_e\cdot {\nabla})	{\bf v}_e = -e n_e ({\bf E}+\frac{1}{c} {\bf v}_e \times {\bf B}) - \nabla p_e + m_e n_e {\bf g} \label{1}
\end{equation}

\begin{equation}
m_i n_i (\partial_t + {\bf v}_i\cdot {\nabla})	{\bf v}_i = e n_i ({\bf E}+\frac{1}{c} {\bf v}_i \times {\bf B}) - \nabla p_i + m_i n_i {\bf g} \label{2}
\end{equation}
where all the physical quantities have standard definitions.
Continuity equations for electrons and ions are,
\begin{equation}
\partial_t n_j + \nabla \cdot (n_j {\bf v}_j)=0 \label{3}
\end{equation}
We assume that the plasma has time-independent gradients of density and temperatures at $t=0$ and evolves with time creating electromagnetic and flow fields. The pressure gradient forces on right hand sides of (\ref{1}) and (\ref{2}) are the source terms for the generation of vorticities and magnetic field. We will find out a 3-D exact analytical solution of these equations which can be used to explain a number of plasma phenomena without dropping any term in (\ref{1}) and (\ref{2}). In addition to it, we will also show that in the case of two dimensional plasma, the  baroclinic vectors $[\nabla n \times (\nabla T_i +\nabla T_e)]$ will produce acceleration to the ions in the direction perpendicular to the 2-D plasma disk and the electron baroclinic vector will be responsible for generating the magnetic field $(\nabla n \times \nabla T_e)$. Both of these phenomena are coupled. The flow velocities of electrons and ions are also coupled through Amperes' law. 

Before discussing the 3-D and 2-D exact analytical solutions of (\ref{1}) and (\ref{2}), it seems important to describe the Biermann battery mechanism which is widely used to investigate the seed magnetic field generation in stars and galaxies as well as in laser produced classical plasmas using two fluid equations. The same mechnism is also described by using the equations of single fluid magnetohydrodynamics (MHD) which will be discussed in section \ref{MHD}.

In 1950, Biermann proposed a mechanism for seed magnetic field generation in stars assuming the plasma has time-independent non-uniform profiles of density and temperatures at time $t=0$ and magnetic field is produced during $t:0 \rightarrow \tau$. It was assumed that $\tau$ lies in between the ion and electron oscillation time periods i.e.  $\omega_{pe}^{-1}<<\tau<<\omega_{pi}^{-1}$ where $\omega_{pj}=(\frac{4 \pi n_j e^2}{m_j})^{\frac{1}{2}}$ is the plasma oscillation frequency of the j-th species. 
Under this approximation, the electron inertia is ignored and ions are assumed to be static. The forces on the right hand side of (\ref{2}) are assumed to balance each other to maintain ions flow to zero. In the limit $m_e \rightarrow 0$, the electron equation of motion becomes,
\begin{equation}
		0=-e n_e {\bf E} - \nabla p_e \label{4}
\end{equation}
The term $\frac{1}{c} e n_e ({\bf v}_e \times {\bf B})$ is dropped in the Biermann battery mechanism keeping in view that there is no magnetic field at $t=0$. 
Using the classical equation of state for electrons pressure, $p_e=n_e T_e$
along with Faraday's law, the curl of (\ref{4}) yields,
\begin{equation}
\partial_t {\bf B} = \frac{c}{e}(\frac{\nabla T_e \times \nabla n_e}{n_e}) \label{5}
\end{equation}
Integrating for $t: 0 \rightarrow \tau$, one obtains,
\begin{equation}
{\bf B}(t)=\frac{c}{e}(\frac{\nabla T_e \times \nabla n_e}{n_e}) \tau \label{6}
\end{equation}
Using (\ref{6}), the magnitude of seed magnetic field in galaxies has been estimated as follows. Let the galactic seed field be $\Delta B$, generation time $\Delta t$, thickness of the clump of gas cloud $h=10^2 pc$ and galaxy radius $R=10^4 pc$. Here $pc$ denotes parsec =$(3.09) \times 10^{13} km$. Then (\ref{6}) can be expressed as (Lazarian1992),
\begin{equation}
\Delta B = \frac{c k_{\beta}}{eS} (\Delta T_e \Delta t) \label{7}
\end{equation}
where $S=h \times R$ the area of the cloud, $\Delta T_e$ is the temperature difference from one end to another end of the gas clump and $k_{\beta}$ is the Boltzmann constant. The author used $\Delta T_e \simeq 10^{6} K$ and scale length of exponential density variation to be $\mid \frac{\nabla n}{n} \mid = h^{-1}$. In (\ref{7}), the temperature has units in Kelvin degrees. Approximating $\nabla T_e= \frac{\Delta T_e}{R}$ and $\tau=10^{9} yrs$, (\ref{7}) gives $\mid B \mid \simeq 3 \times 10^{-17} G$ (Lazarian 1992).

Long ago in \cite{brueckner1974laser}, (\ref{4}) was also applied to estimate the magnetic field generation in initial laser-plasma experiments. Assuming $\nabla n_e = \hat {x}\mid \frac{dn_e}{dx} \mid$, $\nabla T_e = \hat{y} \mid \frac{dT_e}{dx} \mid$, and the scale lengths of gradients to be $L_n=\mid \frac{dn_e}{n_e dx} \mid^{-1}$ and $L_T=\mid \frac{dT_e}{T_e dx} \mid^{-1}$, (\ref{6}) can be written as,
\begin{equation}
{\bf B} = \{ \frac{c}{e} (\frac{T_e}{L_n L_T}) \tau \} \hat{z} \label{8}
\end{equation}
Here $\hat{x}, \hat{y}, \hat{z}$ are unit vectors.
In the initial classical laser plasma experiments, the parameters were $T_e=1 eV$, $n_e = 10^{20} cm^{-3}$, and ion sound speed $c_s=3 \times 10^{7} cm/s$. Then assuming $L_n \approx L_T \approx 0.005 cm$, the magnitude of magnetic field turned out to be $\mid B \mid \approx 0.6 \times 10^{6} G$  which was in agreement with the observations \cite{brueckner1974laser}.

Note that in the experiment $\omega_{pi}^{-1} \simeq 10^{-13} s$ while $\tau = (1.66)\times 10^{-10}s$. This is a contradiction to the initial assumption that $\tau < \omega_{pi}^{-1}$ . Furthermore $\tau$ has been used to be $10^{9} yrs$ in estimating the magnitude of seed magnetic field in galaxies. Considering the above mentioned weaknesses in the application of Biermann battery mechanism, it was suggested that ion dynamics must be included in the theory of magnetic field generation \cite{saleem1996theory,saleem2010nonequilibrium}.

In the present theoretical model, we will keep all the terms and find out an exact 3-D analytical solution of equations (\ref{1}-\ref{3}) with the help of Maxwell's equations. The baroclinic vectors $\nabla \psi \times \nabla T_j$ become the source terms for the generation of magnetic field and vorticities. The interesting point is that all the nonlinear terms vanish due to the chosen spatial profiles of $\psi$ and $T_j$. In case of 2-D plasma, the baroclinic vectors produce acceleration in the direction perpendicular to the plasma disk and give rise to a jet-like flow. The gradients of density and temperatures create time-dependent magnetic ${\bf B}$ and flow fields ${\bf v}_j$.  
We assume longitudinally uniform flows of electrons and ions $\nabla \cdot {\bf v}_j=0$ along with $\partial_t n =0$. Quasi-neutrality  $n_e \simeq n_i=n$ is used under assumption $\mid \lambda_{De}^2 \nabla^2 \mid <<1$, and electron inertia is ignored under the assumption $\mid \partial_t \mid << \omega_{pe}, \Omega_e, \mid c \nabla \mid$. The dynamics of the two fluid plasma can be described by taking curls of electron and ion equations of motion expressed, respectively, as \cite{saleem2010nonequilibrium},
\begin{equation}\label{9}
	\partial_t {\bf B}=\nabla \times ({\bf v}_i \times
	{\bf B})-(\frac{c}{4\pi n e}) \{\nabla \times [(\nabla \times {\bf
		B})\times {\bf B}]\} +\frac{c}{4 \pi n e} \{\nabla \psi\times [(\nabla \times {\bf
	B})\times {\bf B}]\}-\frac{c}{e} (\nabla\psi \times \nabla T_e)
\end{equation}
and
\begin{equation}\label{10}
	\frac{e}{m_i c} \partial_t {\bf B}+\partial_t(\nabla \times{\bf v}_i)=\nabla
	\times [{\bf v}_i\times(\nabla \times{\bf v}_i)]+ 
	\frac{e}{m_i c} \nabla\times({\bf
		v}_i\times{\bf B})+
		\frac{1}{m_i}(\nabla \psi \times \nabla
	T_i)
\end{equation}
where $\psi=\ln \bar{n}$, $\bar{n}=\frac{n}{N_0}$ and $N_0$ is the
arbitrary constant density. The equation of state for ideal classical gas is $p_j=n_j T_j$ \textbf{where temperature $T_j$ has units of energy}. Also the mass conservation requires, $\nabla \psi \cdot {\bf v}_j =0$ and the Amperes' law gives us,
\begin{equation}
{\bf v}_e ={\bf v}_i - \frac{c}{4 \pi e} (\frac{\nabla \times {\bf B}}{n}) \label{12}
\end{equation}
Electron velocity is coupled with the ions velocity and also depends upon curl of magnetic field and the density in (\ref{12}). Thus for longitudinally uniform flows, we also require,
\begin{equation}
\nabla \psi \cdot (\nabla \times {\bf B})=0 \label{13}
\end{equation}
If all nonlinear terms of (\ref{9}) and (\ref{10}) vanish then they reduce, respectively, to,
\begin{equation}
	\partial_t {\bf B}=-\frac{c}{e} (\nabla \psi \times \nabla T_e) \label{14}
\end{equation}
and
\begin{equation}
	\frac{e}{m_i c} \partial_t {\bf B}+\partial_t(\nabla \times {\bf
		v}_i)=\frac{1}{m_i}(\nabla \psi \times \nabla T_i)\label{15}
\end{equation}
The (\ref{14}) and (\ref{15}) give an expression for the generation of ion vorticity by baro clinic vectors,

\begin{equation} 
\partial_t (\nabla \times {\bf v}_i)= \frac{1}{m_i} \nabla \psi \times (\nabla T_{e} + \nabla T_{i}) \label{16}
\end{equation}
Equation (\ref{15}) indicates that ${\bf B}$ and $\nabla \times {\bf v}_i$ can be parallel to each other and our solutions will show that this is true. Therefore, for mathematical convenience, we assume,
\begin{equation} 
{\bf B} = b_0 (\nabla \times {\bf v}_i) \label{17}	
\end{equation}
where $b_0$ is a constant.
Let us look for an exact 3-D analytical solution of the two fluid plasma equations. If we find a suitable function ${\bf v}_i(x,y,z,t)$ which gives,
\begin{equation} 
\nabla \times [{\bf v}_i \times (\nabla \times {\bf v}_i)]=0 \label{18}	
\end{equation}
then due to (\ref{17}), the following expression must also be true,
\begin{equation} 
\nabla \times [{\bf v}_i \times {\bf B}]=0 \label{19}	
\end{equation}
The (\ref{14}) and (\ref{15}) are valid if in addition to (\ref{18}) and (\ref{19}), the following conditions also hold,
\begin{equation} 
\nabla \times [{\bf B} \times (\nabla \times {\bf B})]=0 \label{20}
\end{equation}
and 
\begin{equation} 
	\nabla \psi \times [{\bf B} \times (\nabla \times {\bf B})]=0 \label{21}
\end{equation}
Conditions (\ref{20}) and (\ref{21}) are satisfied if,
\begin{equation} 
\nabla^2 {\bf v}_i = \eta {\bf v}_i \label{22}
\end{equation}
where $\eta$ is a constant. 

Interestingly all the conditions on plasma fields (\ref{18}-\ref{22}) are satisfied, if $\psi(x,y,z)$ and $T_j(x,y,z)$ have following spatial dependence on $(x,y,z)$ coordinates \textbf{(see Appendix A)},

\begin{equation}
\psi(x,y,z)=\psi_0 [e^{\xi(-y+z)}+e^{\xi(-x+z)}+e^{\xi(-x+y)}] \label{23}
\end{equation}

\begin{equation}
T_j(x,y,z)=T_{0j}^{\prime}	(x+y+z) + T_{00j} \label{24}
\end{equation}
where $T_{0j}^{\prime}= \frac{dT_{0j}}{dx}=\frac{dT_{0j}}{dy} =\frac{dT_{0j}}{dz}$, $\psi_0$, $T_{00j}$ and $\xi$ are assumed to be constant. 
Let us define the time evolution of the system by function,  
\begin{equation}
f=C_0 f(t)=C_0 t \label{25}
\end{equation}
where $C_0$ is a constant and we use $C_0=1$ for simplicity.
Then ion velocity is determined by (\ref{16}) as function of $(x,y,z,t)$ \textbf{(see Appendix \ref{appendixA})}.
\begin{eqnarray}
{\bf v}_i(x,y,z,t) &=& a_0 \psi_0  [e^{\xi(-y+z)}+e^{\xi(-x+z)}+e^{\xi(-x+y)}] (1,1,1) t \label{26} \nonumber \\
&= &{\bf a}t = a_0(\psi, \psi, \psi)t
\end{eqnarray}
and magnetic field is given by (\ref{14}),
\begin{eqnarray}
	{\bf B}(x,y,z,t) &=& -(\frac{c}{e} \xi)T_{0e}^{\prime} \psi_0  [(-2 e^{\xi(-y+z)}+e^{\xi(-x+y)}-e^{\xi(-x+z)}),
(2 e^{\xi(-x+z)}+e^{\xi(-y+z)}+e^{\xi(-x+y)}),  \\ \label{27}
&&(-2 e^{\xi(-x+y)}-e^{\xi(-x+z)}+e^{\xi(-y+z)}) ]  f(t) \nonumber
\end{eqnarray}
where $a_0= (\frac{T_{e0}^{\prime} + T_{i0}^{\prime}}{m_i})$ is constant. Note that $\nabla \psi \cdot (\nabla \times {\bf B})=0$ and $\nabla \cdot {\bf B}=0$ are also satisfied. Thus we notice that the highly nonlinear set of coupled partial differential equations is solved exactly if density function $\psi$ and temperatures $T_j$ have the spatial profiles given by (\ref{23}) and (\ref{24}), respectively.

\subsection{Neutral Fluid}
Since ideal neutral fluid dynamics are governed by a simpler set of partial differential equations, therefore a more general form of density function gives an exact 3-D solution. Let us consider the following equations,
 \begin{equation}
\rho(\partial_t + {\bf v} \cdot \nabla) {\bf v} = - \nabla p \label{28}
\end{equation}
and
\begin{equation}
\partial_t \rho + \nabla \cdot (\rho {\bf v})=0 \label{29}
\end{equation}
where $\rho=mn$, $n$ is number density and $m$ is mass of the fluid particle.
We again assume the flow to be longitudinally uniform $\nabla \cdot {\bf v}=0$
with $\partial_t \rho =0$ and define the density function $\psi = \ln \bar{\rho}$ where  $\bar{\rho}=\frac{\rho}{\rho_0}$ and $\rho_0$ is some constant density. The Curl of (\ref{28}) is,

\begin{equation}
\partial_t (\nabla \times {\bf v}) - [\nabla \times {{\bf v} \times (\nabla \times {\bf v})}]=\frac{1}{m} (\nabla \psi \times \nabla T) \label{30}
\end{equation}
The condition $\partial_t \rho =0$ also demands $\nabla \psi \cdot {\bf v}=0$.
Let $\alpha = (-\lambda_1 x - \mu_1 y + \nu_1 z$), $\beta = (-\lambda_2 x - \mu_2 y + \nu_2 z$), and $\gamma = (-\lambda_3 x - \mu_3 y + \nu_3 z$) where $\nu_k = \lambda_k + \mu_k$ and $k =1,2,3$. If $\psi=\psi (x,y,z)$ and $T=T(x,y,z)$ have following forms in 3-D space, 
\begin{equation}
\psi(x,y,z) = \psi_0 (e^{\alpha}+e^{\beta}  +e^{\gamma}) \label{31}
\end{equation}

\begin{equation}
T(x,y,z)= T_0^{\prime}(x+y+z) + T_{00} \label{32}
\end{equation}
where $T_0^{\prime}$ and $T_{00}$ are constants, then the nonlinear term of (\ref{30}) vanishes \textbf{(see Appendix \ref{appendixB})} and it reduces to,
\begin{equation}
\partial_t (\nabla \times {\bf v})=\frac{1}{m} (\nabla \psi \times \nabla T) \label{33}
\end{equation}
Here $\psi_0, \lambda_k, \mu_k, \nu_k$ and $T_0^{\prime}= \frac{dT_0}{dx}=\frac{dT_0}{dy} =\frac{dT_0}{dz} $ are constants.
The following form of velocity field is generated, 
\begin{equation}
{\bf v}(x,y,z,t)=a_0 \psi (1,1,1) f(t) = {\bf a} t \label{34}
\end{equation}
which satisfies (\ref{33}) where $a_0=\frac{T^{\prime}}{m}$ is constant. 

\subsection{Magnetohydrodynamics (MHD)}\label{MHD}
Momentum conservation equation for single fluid plasma, the magnetohydrodynamics (MHD), is,
\begin{equation}
\rho (\partial_t + {\bf v}\cdot {\nabla})	{\bf v} = \frac{1}{c}{\bf j}\times{\bf B} - \nabla p + \rho {\bf g} \label{35}
\end{equation}
Here ${\bf v}$ is the bulk plasma velocity, $\rho= n m_i $ and $p=p_e+p_i$. Genralized Ohm's law is,
\begin{equation}
{\bf E} + \frac{1}{c}{\bf v}\times{\bf B}= \frac{1}{\sigma} {\bf j} + \frac{1}{e n}(\frac{1}{c}{\bf j}\times{\bf B} - \nabla p_e) \label{36}
\end{equation}
Assuming collision-less plasma, the first term on rhs of (\ref{36}) is ignored. Biermann battery mechanism for the generation of magnetic field can also be considered using MHD equations (\cite{boyd2003physics}). The terms containing ${\bf B}$ are neglected assuming ${\bf B} =0$ at $t=0$. Plasma bulk motion is also neglected assuming that the forces on right hand side of (\ref{35}) balance each other. Then Ohm's law (\ref{36}) reduces to (\ref{4}) which has been obtained through the two fluid plasma equations.
But Biermann mechanism does not take into account the creation of plasma vorticity. Here we show that the MHD equations also predict the generation of bulk plasma flow with non-zero vorticity $\nabla \times {\bf v} \neq 0$. Curl of (\ref{35}) gives,
\begin{equation}
	\partial_t ({\nabla \times \bf v}) - \nabla
	\times [{\bf v}\times(\nabla \times{\bf v})]-\frac{1}{4 \pi \rho}
	\nabla \times [(\nabla \times {\bf B}) \times{\bf B} ]  \label{37}+\frac{1}{4 \pi \rho} \nabla \psi \times [(\nabla \times {\bf B}) \times{\bf B}]= \frac{1}{\rho} (\nabla \psi \times \nabla p) \label{37}
\end{equation}
Nonlinear terms in eq.(\ref{37}) vanish if $\psi$ is given by (\ref{23}) and $T$ is given by (\ref{32}). Then (\ref{37}) reduces to,
\begin{equation}\label{38}
	\partial_t ({\nabla \times \bf v}) = \frac{1}{\rho}(\nabla \psi \times \nabla p)
\end{equation}
If we use the form of $\psi$ defined by (\ref{23}) along with temperature profile given by (\ref{32}) then (\ref{38}) yields the following form of plasma flow,

\begin{equation}
{\bf v}(x,y,z,t)=(\frac{T_0^{\prime}}{m_i}) \psi_0 [e^{\xi(-y+z)}+e^{\xi(-x+z)}+e^{\xi(-x+y)}] (1,1,1) t \label{39}
\end{equation}
Thus we conclude that magnetic field and plasma vorticity are generated by the baro clinic vector simultaneously in single fluid plasma model as well.

\section{Creation of Solar Spicules}\label{solarspicule}
Plasma in these structures has the predominant flow in upward direction perpendicular to the surface.
For $\alpha = (\mu x + \nu y)$, $\beta=(\mu x - \nu y)$ and neglecting the $\gamma$ term, the $\psi =\psi(x,y)$ function becomes independent of the z-coordinate,
\begin{equation}
\psi(x,y) = \psi_0 (e^{\mu x + \nu y}+ e^{\mu x - \nu y}) \label{40}
\end{equation}

\begin{figure*}
\gridline{\fig{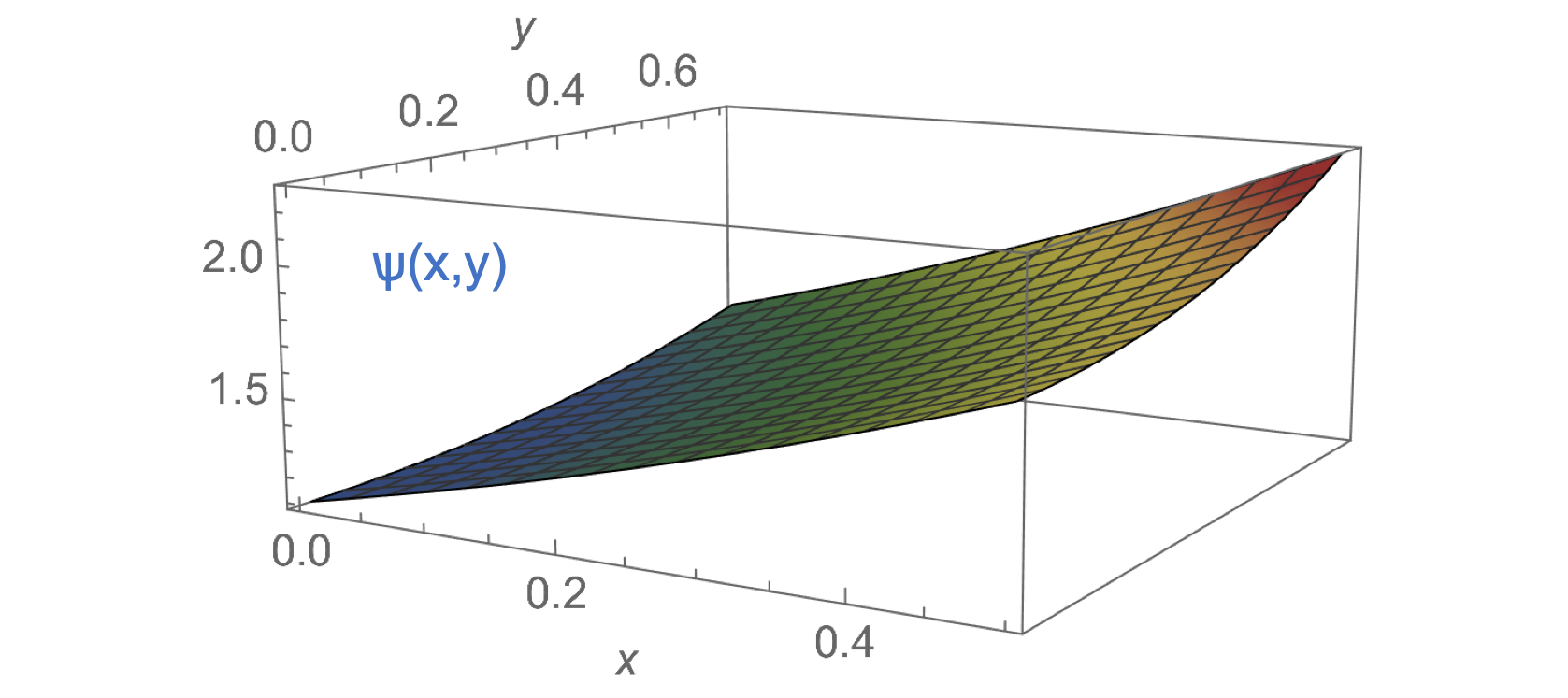}{0.3\textwidth}{(a)}
          \fig{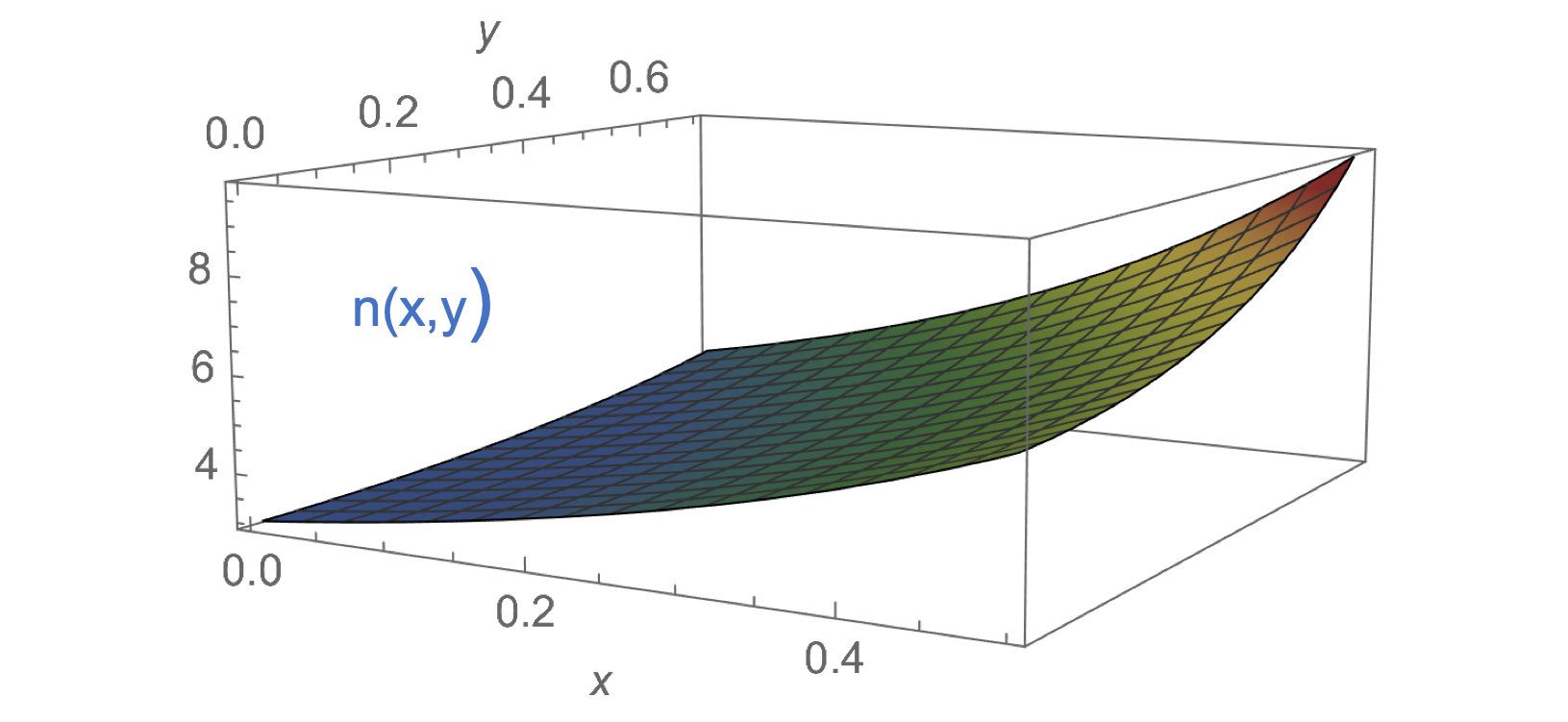}{0.3\textwidth}{(b)}
          }
\caption{(a) The profile of $\psi(x,y)$. (b) The profile of $n(x,y)$..
\label{pyramid}}
\end{figure*}
Let us assume that temperatures $T_j$ vary linearly only along z-axis as,
\begin{equation}
T_j(z) = T_{0j}^{\prime} (z) + T_{00j} \label{41}
\end{equation}
where $T_{00j}$ and $T_{0j}^{\prime} = \frac{d T_{0j}}{dz}$ are  constants. 
Use of above profiles of $\psi$ and $T_j$ in (\ref{14}) yields,
\begin{equation}
{\bf B}(x,y) = - \frac{c}{e} T_{0e}^{\prime} (\partial_y \psi, - \partial_x \psi, 0) f(t) = {\bf B}_{g} f(t)\label{42}
\end{equation}
where ${\bf B}_{g} = {\bf B}(x,y)$ is the spatial part of the generated field which has following form,

\begin{equation}\label{43}
{\bf B}_{g}(x,y) = \frac{cT_{e0}^{\prime}}{e}\mu \psi_0 [- \frac{\nu}{\mu}(e^{\mu x - \nu y}- e^{\mu x - \nu y}), (e^{\mu x + \nu y}+ e^{\mu x - \nu y}), 0]
\end{equation}
Then (\ref{16}) gives ion plasma vorticity, 
\begin{equation}\label{44}
\partial_t (\nabla \times {\bf v}_i) = a_0 \psi_0 \mu[\frac{\nu}{\mu}(e^{\mu x - \nu y}- e^{\mu x - \nu y}), - (e^{\mu x + \nu y} + e^{\mu x - \nu y}), 0] 
\end{equation}
Here $a_0$ has the same definition as has been mentioned below (26). Equation (43) determines the unidirectional flow in z-direction perpendicular to the plasma surface given by,
\begin{equation}
{\bf v}_i(x,y)=	[(a_0) \psi(x,y) \hat{z}] f(t) = {\bf a} f(t)={\bf a} t \label{45}
\end{equation}
where ${\bf a}(x,y)=a_0 \psi(x,y) \hat{z}$ is the upward ions acceleration. 
Let us assume that the spicules are formed by many small slabs created one by one due to gradients of density and temperatures at bottom of these structures. These slabs move upward due to the acceleration produced by gradients and as they reach upper regions, the temperature gradient approaches zero and velocity becomes constant as observed. 
\begin{figure}[htb]
    \centering
    \includegraphics[width = 2.5in, height = 2.5 in]{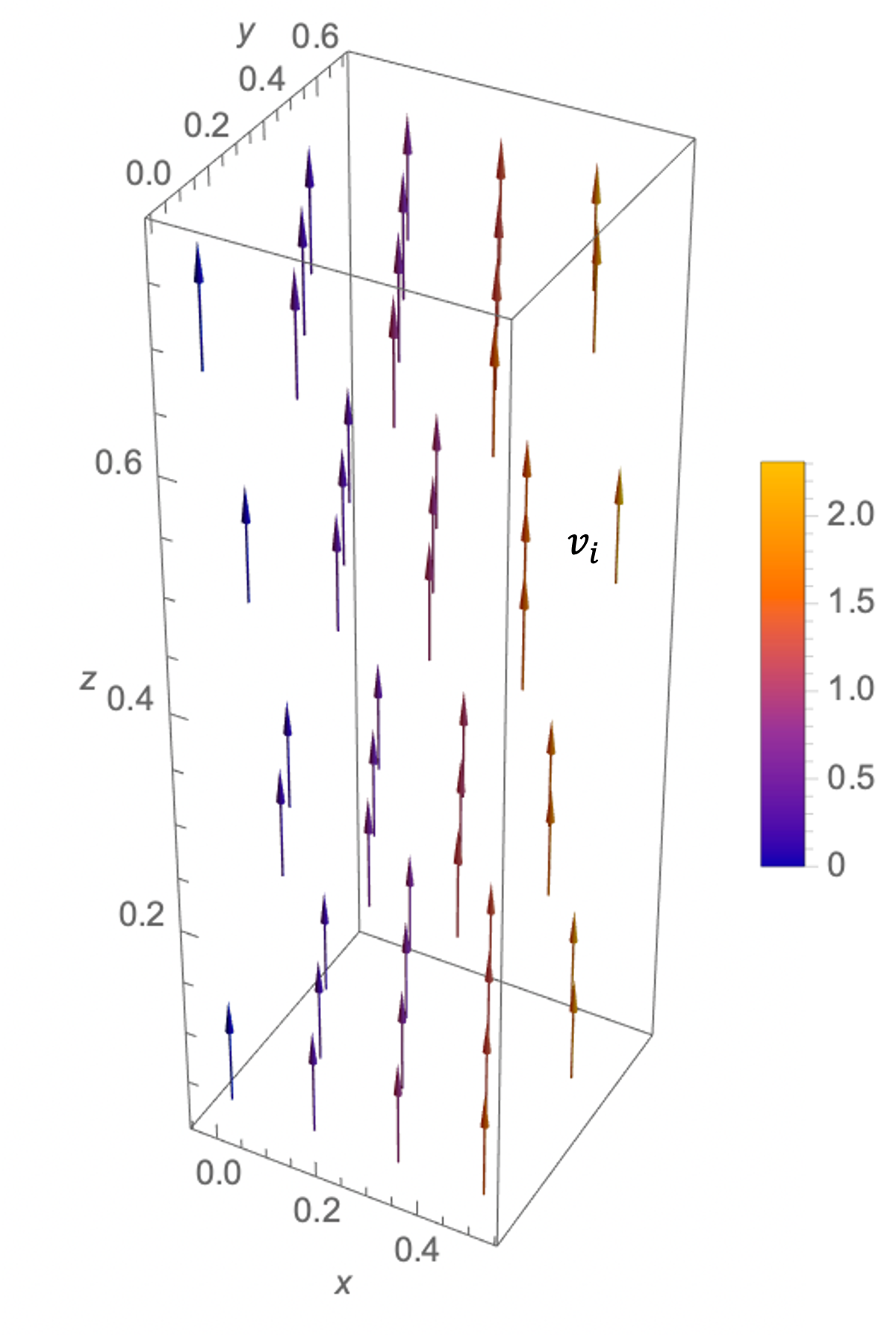}
    \caption{Velocity field $\bf{v}_i$ in a solar spicule when $T_i$ is a function of the z- coordinate only and $\psi=\psi(x,y)$.}
    \label{velocity}
\end{figure}
Since spicules have strong ambient magnetic field as well, therefore we add a constant magnetic field to the weak magnetic field generated by the baro clinic vectors and write the total field within the structures ${\bf B}_{st}$ as,
\begin{equation}
{\bf B}_{st}={\bf B}_g(x,y,t)+{\bf B}_0 \label{46}
\end{equation}
The ${\bf B}_0=B_0 \hat{z}$ is the constant external magnetic field produced at the footprints of the structures by physical mechanisms operative at the solar surface or below. Note that the 2-D theoretical model remains valid when we express the total magnetic field in the form of (\ref{46}).

\begin{figure}[htb]
    \centering
    \includegraphics[width = 2.5in, height = 2.5 in]{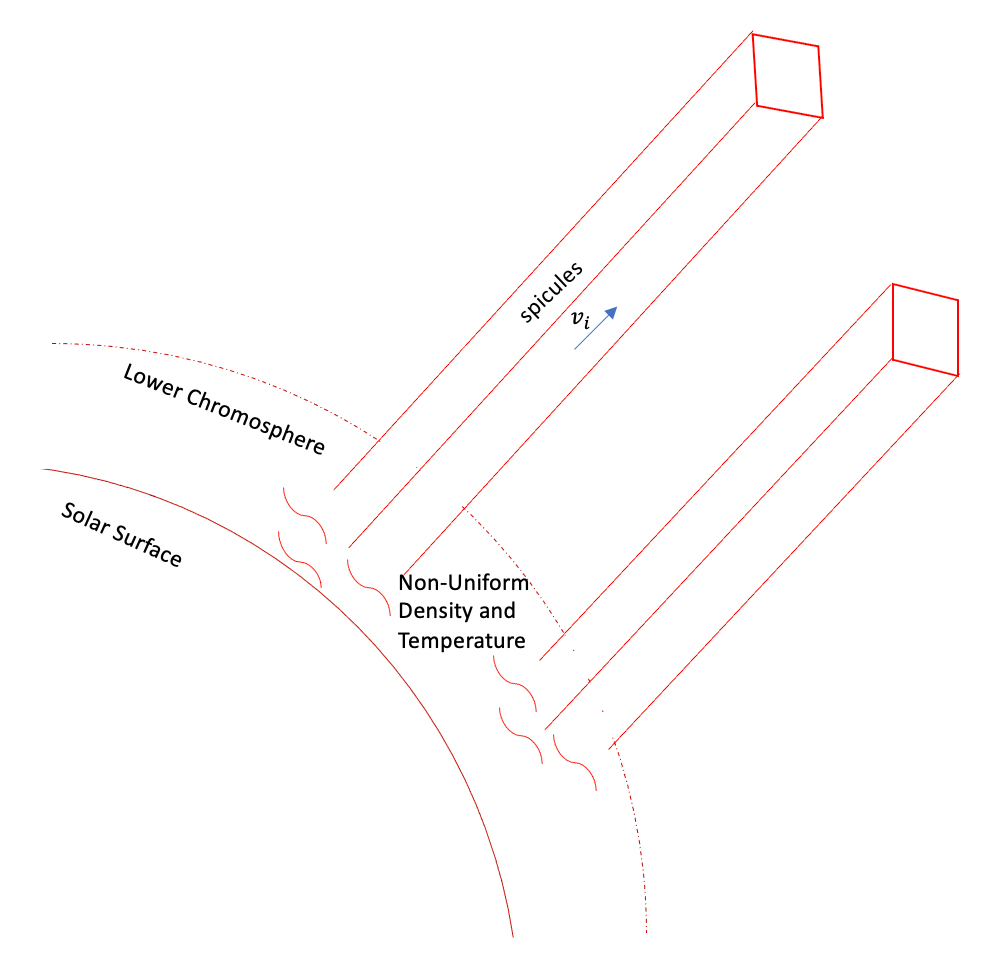}
    \caption{Schematic diagram of accelerated plasma moving upward due to $ \nabla \psi \times \nabla T_i \neq 0$. }
    \label{my_label}
\end{figure}

We will show now that for suitable choice of numerical values in our model, the acceleration of the plasma can be larger than the solar surface gravity and that the height of the solar spicules turns out to be closer to the heights observed. For the plasma slab we assume $x: 0 \rightarrow x_m$, $y: 0 \rightarrow y_m$, and choose height $h=z_2-z_1$ in direction perpendicular to the 2-D plasma in xy-plane. Forms of $\psi(x,y)$ and $n(x,y)$ for our choice of numerical values $\mu x_m=0.5$ and $\nu y_m=0.7$ are shown in Fig.\ref{pyramid}. Let $T_e \simeq T_i=T$ and consider the slab formed at bottom of the spicule 
with dimensions $x_m=3 \times 10^{7} cm$, $y_m= 4 \times 10^{7} cm$ and $h=3 \times 10^7 cm$. We consider $T=3 \times 10^3 K$ in the upper region of solar surface ($z=z_1$) and $T= 5 \times 10^{4} K$ in the lower chromosphere ($z=z_2$). Approximating $T^{\prime}=\frac{(T(z_0)-T(0))}{h}$, we find $a_0=\frac{2 T^{\prime}}{m_i}=(3.3)\times 10^{4} cm/s^2$ and hence $g_{\odot}<a_0$. Velocities at different points can be expressed as ${\bf v}(x,y)= \hat{z}( a(x,y)- g_{\odot}) \tau$ where $\tau$ is evolution time of the inhomogeneous plasma slab. Choosing $\tau= 10 s$, the magnitude of initial velocities turns out to be $v(x_m, y_m)=7.5 \times 10^{4} cm/s$ and $v(0,0)= 3.63 \times 10^{4} cm/s$.  Later the solar attraction produces deceleration and the structures height is limited by this process. If we consider the constant velocity magnitude to be $5 \times 10^5/s$, then during the life time $\tau_L = 300 s$, the spicule attains the height $H= 1.4 \times 10^8 cm$. Magnetic field produced by baro clinic vectors is too small of the order of $10^{-7} G$ compared to the ambient magnetic field $B_0=2 G$ which is produced at the footprints of the structure in solar surface. Form of velocity function in solar spicule is shown in Fig.(\ref{velocity}) and formation of spicules is schematically shown in Fig. (\ref{my_label}). The baro clinic vectors $\nabla \psi \times (\nabla T_i + \nabla T_e)$ act as the source for producing ions acceleration in the upward direction ${\bf a}$. Ions can flow in the upward direction when $\mid {\bf g}_{\odot} \mid < \mid {\bf a} \mid$. Since ions velocity and the curl of the magnetic field are parallel, therefore  (\ref{12}) shows that electrons follow the ions motion along z-axis.

\section{Summary}
The exact 3-D analytical solutions of ideal two fluid plasma, MHD, and neutral fluid equations have been presented using Cartesian geometry. We have assumed exponential like spatial profile for density in 3-D and linear dependence of ion and electron temperatures on $(x,y,z)$ coordinates with $T_i \neq T_e$. The velocity ${\bf v}_j$ and magnetic ${\bf B}_g$ fields evolve  with time and become functions of $(x,y,z,t)$ coordinates.
One can possibly obtain several similar solutions by varying signs of exponential functions in the definition of $\psi$. The formation of solar spicules has been discussed using 2-D version of this model. These structures disappear at higher altitudes where temperature gradient vanishes. 

It is very interesting that similar spicule structures can also be obtained using MHD equation (\ref{38}) or neutral fluid equation (\ref{33}), but in these cases the very weak magnetic field ${\bf B}_g$ will not be associated with the vorticity generation. This indicates that the non-parallel density and temperature gradients $\nabla \psi \times \nabla T_j \neq 0$ mainly produce acceleration in both plasma and neutral fluid. This acceleration turns out to be in the direction perpendicular to the 2-D material when temperatures vary only in vertical direction. The Biermann battery effect has a small contribution of electron baro clinic vector while it creates large acceleration when dynamics of the whole system of two fluid plasma is considered in detail. If $\nabla \times {\bf v}_i=0$ is assumed, then (\ref{14}) and (\ref{15}) show that in this case either plasma attains equilibrium $(T_i=T_e)$ with $\partial_t {\bf B}_g=0$ or $\nabla T_e= - \nabla T_i$. The forms of $\psi$ and $T_j$ given in (\ref{15}) and (\ref{16}), respectively, are also applicable to ideal magnetohydrodynamics (MHD). 

The 3-D and 2-D models discussed above can be applied to explain formation of several structures emerging in natural and laboratory environments. Formation of solar spicules, solar coronal loops and prominences can all be explained on the basis of this theoretical model. Furthermore, this model is also useful for understanding the mechanisms of coronal mass ejection (CME) and formation of jets ejected from astrophysical objects consisting of classical plasmas and neutral fluids. 

\appendix

\section{\textbf{Two fluid plasma: $\nabla \times [{\bf v}_i \times (\nabla \times {\bf v}_i)]=0$}}\label{appendixA}
Here we show that with the spatial functions $\psi$ and $T_j$ defined in (\ref{23}) and (\ref{24}), the ions velocity and magnetic field are forced to have the forms given in (\ref{26}) and (\ref{27}), respectively and consequently the nonlinear term $\nabla \times [{\bf v}_i \times (\nabla \times {\bf v}_i)]$ vanishes. To write equations in simpler form, let us define $S_1=(-y+z)$, $S_2=(-x+z)$ and $S_3=(-x+y)$. Then the function $\psi$ can be expressed as,

\begin{equation}
\psi = \psi_0 [e^{\xi S_1} + e^{\xi S_2} + e^{\xi S_3}] \label{A-1}
\end{equation}
The cross product of $\nabla \psi$ and $\nabla T_j$ becomes,
\begin{equation}
(\nabla \psi \times \nabla T_j) = T_{0j}^{\prime}[(\partial_y \psi - \partial_z \psi ), (\partial_z \psi - \partial_x \psi ), (\partial_x \psi - \partial_y \psi )]
\end{equation}
or
\begin{equation}
(\nabla \psi \times \nabla T_j) = (\xi \psi_0 T_{0j}^{\prime})[(-2e^{\xi S_1}+ e^{\xi S_3} - e^{\xi S_2}),(2e^{\xi S_2}+ e^{\xi S_1} + e^{\xi S_3}), (-2e^{\xi S_3}- e^{\xi S_2} + e^{\xi S_1})] 
\end{equation}
Using (A3) in (\ref{14}), we obtain,

\begin{equation}
{\bf B}=(- \frac{c}{e}\xi \psi_0 T_{0e}^{\prime})[(-2e^{\xi S_1}+ e^{\xi S_3} - e^{\xi S_2}),(2e^{\xi S_2}+ e^{\xi S_1} + e^{\xi S_3}), 
(-2e^{\xi S_3}- e^{\xi S_2} + e^{\xi S_1})] f(t)
\end{equation}
If ions flow velocity is expressed as, ${\bf v}_i = a_0 (\psi \hat{x} + \psi \hat{y} + \psi \hat{z}) f(t)$ then one can easily show that $\nabla \cdot {\bf v}_i=0$. Ions vorticity becomes,
\begin{equation}
(\nabla \times {\bf v}_i) = a_0 [(\partial_y \psi - \partial_z \psi ), (\partial_z \psi - \partial_x \psi ), (\partial_x \psi - \partial_y \psi )] f(t) \label{A5}
\end{equation}
It shows that ions vorticity $(\nabla \times {\bf v}_i)$ is parallel to ${\bf B}$ which is parallel to electron baro clinic vector $(\nabla \psi \times \nabla T_e)$.
Note that,$\nabla^2 \psi= 2 \xi^2 \psi$ which yields, $\nabla^2 {\bf v}_i= 2 \xi^2 {\bf v}_i$ and hence, $\nabla \times {\bf B} = -b_0 (2 \xi^2) {\bf v}_i$. Thus ${\bf v}_i$ turns out to be parallel to $\nabla \times {\bf B}$. Equation (\ref{23}) yields,
\begin{equation}
\psi^2= \psi_0^2 [(e^{ 2\xi S_1} + e^{2\xi S_2} + e^{2\xi S_3}) + 2 (e^{ \xi (S_1+S_2)} + e^{\xi S_2} + e^{\xi (S_2+S_3)})]
\end{equation}
which gives,
\begin{eqnarray}\label{A11}
[{\bf v}_i \times (\nabla \times {\bf v}_i)]&=& \frac{1}{2}(a_0 t)^2 [(2\partial_x \psi^2-\partial_y \psi^2- \partial_z \psi^2), (2\partial_y \psi^2-\partial_x \psi^2- \partial_x \psi^2), 
\\ \nonumber 
&&(2\partial_z \psi^2-\partial_x \psi^2- \partial_y \psi^2)] 
\end{eqnarray}
Using (A6) in (\ref{A11}) and following straightforward but cumbersome algebra, one can easily show that,
$\nabla \times [{\bf v}_i \times (\nabla \times {\bf v}_i)]=0$.

\section{\textbf{Neutral fluid: $\nabla \times [{\bf v}_i \times (\nabla \times {\bf v}_i)]=0$}}\label{appendixB}

Equation (\ref{31}) gives,
\begin{equation} \label{B1}
\nabla \psi= \psi_0 [(-\lambda_1 e^{\alpha} - \lambda_2 e^{\beta} - \lambda_3 e^{\gamma}), (-\mu_1 e^{\alpha} - \mu_2 e^{\beta} - \mu_3 e^{\gamma}), 
{(\nu_1 e^{\alpha} + \nu_2 e^{\beta} + \nu_3 e^{\gamma})}]
\end{equation}
Equations (\ref{B1}) and (\ref{32}) yield,
\begin{eqnarray}
(\nabla \psi \times \nabla T) &=& \psi_0 T_0^{\prime}[{-(\mu_1 + \nu_1)e^{\alpha} -(\mu_2 + \nu_2)e^{\beta} -(\mu_3 + \nu_3)e^{\gamma}},
{(\nu_1 + \lambda_1)e^{\alpha} +(\nu_2 + \lambda_2)e^{\beta} +(\nu_3 + \lambda_3)e^{\gamma}}, 
\\ \nonumber 
&&
{(\mu_1 - \lambda_1)e^{\alpha} + (\mu_2 - \lambda_2)e^{\beta} +(\mu_3 - \lambda_3)e^{\gamma}}]
\end{eqnarray}
Then using (\ref{34}), we obtain,
\begin{equation}
\nabla \cdot {\bf v}= a_0 f(t) \psi_0 [(-\lambda_1-\mu_1+\nu_1)e^{\alpha} + (-\lambda_2-\mu_2+\nu_2)e^{\beta}) + (-\lambda_3-\mu_3+\nu_3)e^{\gamma}]
\end{equation}
and 
\begin{equation}
\nabla \psi \cdot {\bf v}= \psi_0 a_0 f(t) (e^{\alpha} + e^{\beta} + e^{\gamma})[(\nu_1-\lambda_1-\mu_1) e^{\alpha}+ (\nu_2-\lambda_2-\mu_2) e^{\beta}+ (\nu_3-\lambda_3-\mu_3) e^{\gamma}]
\end{equation}
For $\nu_k = \lambda_k + \mu_k$ where $k=1,2,3$, we have $\nabla \cdot {\bf v}=0$ and $\nabla \psi \cdot {\bf v}=0$. Form of velocity (\ref{34}) gives,
\begin{eqnarray}
\nabla \times {\bf v} &=&	a_0 f(t)[\{-(\mu_1 + \nu_1)e^{\alpha} -(\mu_2 + \nu_2)e^{\beta} -(\mu_3 + \nu_3)e^{\gamma}\}, 
\{(\nu_1 + \lambda_1)e^{\alpha} +(\nu_2 + \lambda_2)e^{\beta} +(\nu_3 + \lambda_3)e^{\gamma}\}, 
\\ \nonumber 
&&
\{(\mu_1 - \lambda_1)e^{\alpha} +(\mu_2 - \lambda_2)e^{\beta} -(\mu_3 + \lambda_3)e^{\gamma}\}]
\end{eqnarray}

Let $e_1=(e^{2\alpha}+e^{(\alpha + \beta)}+e^{(\alpha + \gamma)})$, $e_2=(e^{2 \beta}+e^{(\alpha + \beta)}+e^{(\beta + \gamma)})$, $e_3=(e^{2 \gamma}+e^{(\alpha + \gamma)}+e^{(\beta + \gamma)})$. Then we can write,

\begin{equation}
{\bf v} \times (\nabla \times {\bf v}) = 3 a_0^2  [(-\lambda_1 e_1 -\lambda_2 e_2- \lambda_3 e_3), (-\mu_1 e_1 -\mu_2 e_2- \mu_3 e_3), 
(\nu_1 e_1 + \nu_2 e_2 + \nu_3 e_3)] (f(t))^2
\end{equation}

Again following a little laborious but straightforward algebra, we obtain from (\ref{34}),
\begin{equation}
\nabla \times [{\bf v} \times (\nabla \times {\bf v})]=0
\end{equation}

\bibliography{bibliography}{}
\bibliographystyle{aasjournal}



\end{document}